\documentclass[twocolumn,showpacs,superscriptaddress,amsmath,amssymb,aps,prl]{revtex4-1}
\UseRawInputEncoding
\usepackage{bm}
\usepackage{graphicx}
\usepackage{dcolumn}
\usepackage{amsmath,txfonts}
\usepackage{epstopdf}
\usepackage[mathlines]{lineno}
\usepackage{color}
\usepackage[colorlinks=true,linkcolor=blue,citecolor=blue]{hyperref}
\usepackage[normalem]{ulem}

\begin{document}

\title{Pokemon:~Protected Logic Qubit Derived from the 0-$\pi$ Qubit}

\author{J. Q. You}
\email{jqyou@zju.edu.cn}
\affiliation{Interdisciplinary Center of Quantum Information, State Key Laboratory of Modern Optical Instrumentation, and Zhejiang Province Key Laboratory of Quantum Technology and Device, Department of Physics, Zhejiang University, Hangzhou 310027, China}

\author{Franco Nori}
\affiliation{Theoretical Quantum Physics Laboratory, RIKEN Cluster for Pioneering Research, Wako-shi, Saitama 351-0198, Japan}
\affiliation{Physics Department, University of Michigan, Ann Arbor, Michigan 48109-1040, USA}

\begin{abstract}
We propose a new protected logic qubit called pokemon, which is derived from the 0-$\pi$ qubit by harnessing one capacitively shunted inductor and two capacitively shunted Josephson junctions embedded in a superconducting loop. Similar to the 0-$\pi$ qubit, the two basis states of the proposed qubit are separated by a high barrier, but their wave functions are highly localized along both axis directions of the two-dimensional parameter space, instead of the highly localized wave functions along only one axis direction in the 0-$\pi$ qubit. This makes the pokemon qubit more protected. For instance, the relaxation of the pokemon qubit is exponentially reduced by two equally important factors, while the relaxation of the 0-$\pi$ qubit is exponentially reduced by only one factor. Moreover, we show that the inductor in the pokemon can be replaced by a nonlinear inductor using, e.g., a pair or two pairs of Josephson junctions. This offers an experimentally promising way to implement next-generation superconducting qubits with even higher quantum coherence.
\end{abstract}

\date{\today}

\maketitle

{\it Introduction.}{\bf---}Quantum computers can outperform their classical counterparts in simulating many-body quantum systems~\cite{Nori-RMP-14,Aspuru-Guzik-RMP-20} and implementing important algorithms~\cite{Nielsen-Chuang}, owing to their exponentially large capacity in storing and processing information. These quantum advantages can be demonstrated when qubits in quantum computers achieve sufficiently high quantum coherence. In recent years, superconducting qubits~\cite{Clarke-PRL-85,Devoret-PScip-98,Nakamura-Nature-1999,Mooij-Sci-99,Vion-Science-02,Yu-Sci-02,Martinis-PRL-02,Nori-PhysRep-17} based on Josephson-junction circuits have indeed been considerably improved in their quantum coherence, showing quantum advantage using tens of superconducting qubits~\cite{Martinis-Nature-19,Zhu-PRL-21}.

This quantum-coherence enhancement of superconducting qubits was achieved by shunting a large capacitance to the small Josephson junction in the circuit to reduce the sensitivity of the qubits to the charge noise~\cite{Siddiqi-NatRev-21,Tsai-JAP-21}. This was proposed in the capacitively shunted flux qubit~\cite{You-PRB-07} and later implemented experimentally~\cite{Steffen-PRL-10,Yan-NC-16}. Also, it was proposed for the capacitively shunted Cooper-pair box, called the {\it transmon}~\cite{Koch-PRA-07}. To be tunable and easily coupled to the neighboring qubits, the transmon was latter modified as the {\it Xmon}~\cite{Martinis-PRL-13} by both replacing the small Josephson junction with a superconducting quantum interference device (SQUID) and connecting a cross-shaped electrode to the SQUID. In addition, the {\it gatemon}~\cite{Larsen-PRL-15}, which is also a transmon-like device, and  the capacitively shunted fluxonium~\cite{Earnest-PRL-18} were implemented. Generally speaking, these high-coherence superconducting qubits are encoded only on {\it one} degree of freedom related to the small Josephson junction (or SQUID) in the circuit, whereas other circuit elements are to adjust the anharmonicity of the qubit~\cite{Tsai-JAP-21,Wendin-Review-17}.

\begin{figure}
\includegraphics[width=0.48\textwidth]{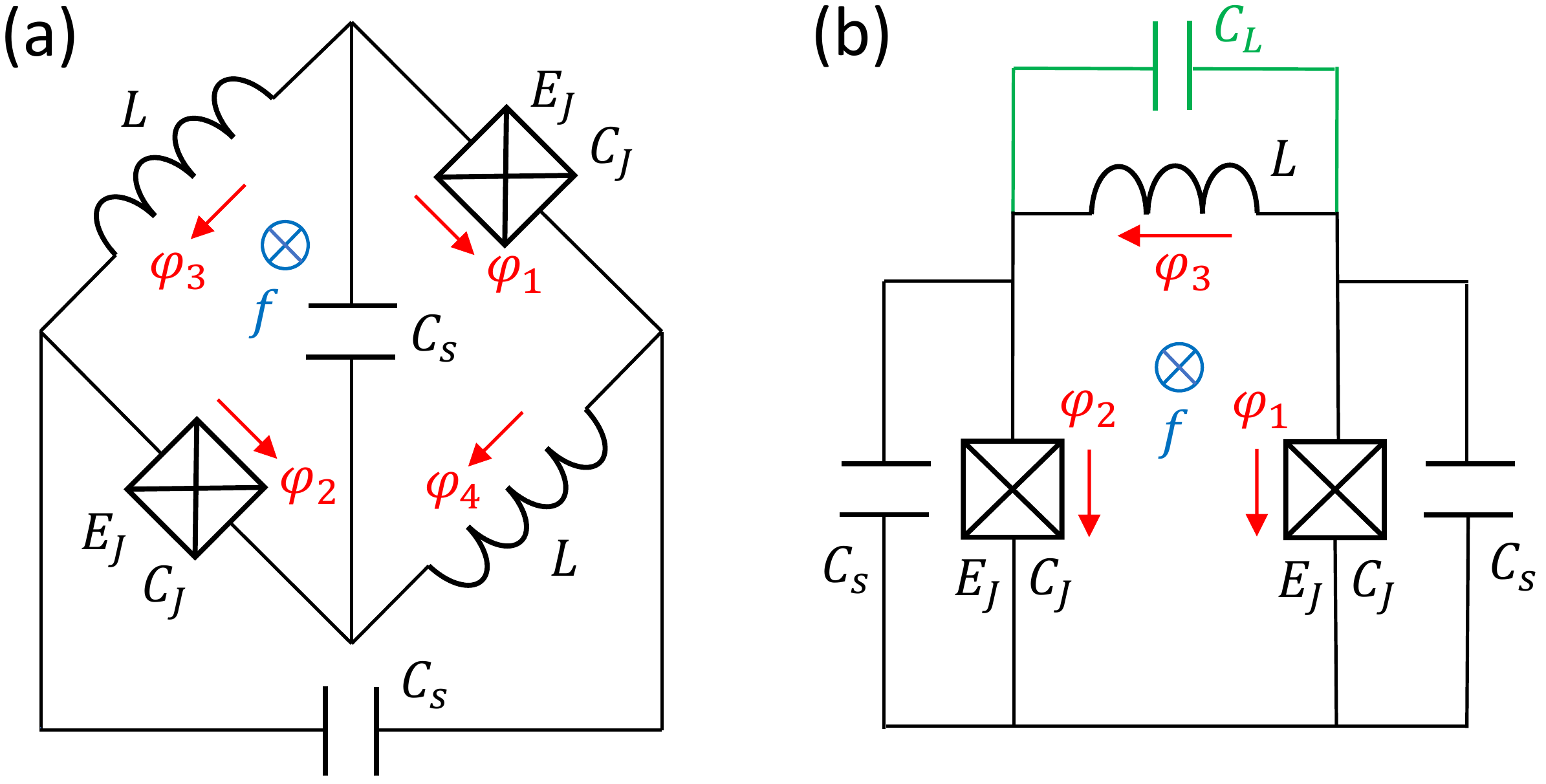}
\caption{(a) The circuit for the 0-$\pi$ qubit in Ref.~\cite{Brooks-PRA-13}, composed of a superconducting loop with two identical Josephson junctions (each has a coupling energy $E_J$ and capacitance $C_J$), two identical inductors $L$, and two intersecting capacitors of the same capacitance $C_s$. (b) The circuit for the protected logic qubit called pokemon, consisting of a superconducting loop with two capacitively shunted Josephson junctions and an inductor $L$ shunted by a capacitance $C_L$. Each junction has a coupling energy $E_J$ and capacitance $C_J$, and is also shunted with a large capacitance $C_s$. Without the (green) capacitance $C_L$, the circuit is reduced to a bifluxon qubit~\cite{bifluxon-20}, but each junction is shunted by a large capacitance. In both (a) and (b), as well as in Fig.~\ref{fig3} below, each (red) arrow indicates the assigned phase-drop direction for the junction or inductor. The reduced magnetic flux threading the loop is $f=\Phi_{\rm ext}/\Phi_0$.}
\label{fig1}
\end{figure}

However, a practical quantum computer should be fault-tolerant, requiring a significantly increased system scale~\cite{Preskill-Quantum-18}. This needs qubits with even higher quantum coherence. A protected logic qubit called the 0-$\pi$ qubit was proposed~\cite{Brooks-PRA-13} and implemented very recently~\cite{Gyenis-PRXQ-21}, which harnesses two Josephson junctions and two inductors embedded in a superconducting loop and shunted with two intersecting capacitors [see Fig.~\ref{fig1}(a)]. This superconducting qubit is encoded on the {\it two} degrees of freedom of the Josephson junctions, with two isolated wave functions in the two-dimensional (2D) parameter space acting as the basis states of the qubit. While these two states are separated by a high barrier, they are highly localized only along one axis direction in the 2D parameter space. Also, the circuitry of the 0-$\pi$ qubit becomes complex when implementing each of the two inductors with a large number of Josephson junctions~\cite{Gyenis-PRXQ-21}.

In this work, we propose a simplified structure to implement a new protected logic qubit, called {\it pokemon}, which is derived from the 0-$\pi$ qubit by using one capacitively shunted inductor and two capacitively shunted Josephson junctions embedded in a superconducting loop. The two basis states of the proposed qubit are also separated by a high barrier. Moreover, they are highly localized along {\it both} axis directions of the 2D parameter space, instead of the highly localized basis states along {\it only one} axis direction in the 0-$\pi$ qubit. These make the pokemon more protected. For instance, the relaxation of the proposed qubit is exponentially reduced by two equally important factors, while the relaxation of the 0-$\pi$ qubit is exponentially reduced by only one factor. Also, we show that the inductor in the pokemon can be replaced by a nonlinear inductor using, e.g., a pair or two pairs of Josephson junctions, instead of using a large number of Josephson junctions to achieve a linear inductor. This provides an experimentally promising way to implement next-generation superconducting qubits with even higher coherence to carry out more demanding tasks in quantum computing.

{\it Pokemon derived from the 0-$\pi$ qubit.}{\bf---}In contrast to the 0-$\pi$ qubit in Fig.~\ref{fig1}(a), the protected logic qubit called pokemon is composed of two (identical) capacitively shunted Josephson junctions and a capacitively shunted inductor embedded in a superconducting loop [see Fig.~\ref{fig1}(b)]. Also, we harness a large shunting capacitance for each Josephson junction to reduce the noise effect on the qubit.

In this pokemon, the phase drops across the two Josephson junctions and the inductor are constrained by the fluxoid quantization condition: $\varphi_1-\varphi_2-\varphi_3+2\pi f=0$, with $f\equiv\Phi_{\rm ext}/\Phi_0$, where $\Phi_{\rm ext}$ is the applied magnetic flux threading the loop and $\Phi_0=h/2e$ is the flux quantum. The voltages across the two Josephson junctions and the inductor are related to the corresponding phase drops by $V_i=(\Phi_0/2\pi)\dot{\varphi}_i$. The electric energy of the pokemon is $T=\frac{1}{2}(C_s+C_J)(V_1^2+V_2^2)+\frac{1}{2}C_LV_3^2=\frac{1}{4}(C_s+C_J)(V_1+V_2)^2+\frac{1}{4}(C_s+C_J+2C_L)V_3^2$.
Using the relationship between the voltage and phase drop, as well as the canonical coordinates $\theta\equiv\frac{1}{2}(\varphi_1+\varphi_2)$ and $\varphi\equiv\frac{1}{2}\varphi_3$, we can express the electric energy as $T=(\Phi_0/2\pi)^2[(C_s+C_J){\dot\theta}^2+(C_s+C_J+2C_L){\dot\varphi}^2]$.

The Lagrangian of the pokemon is ${\cal L}=T-U$, where the potential is
$U=E_J(1-\cos\varphi_1)+E_J(1-\cos\varphi_2)+(\Phi_0/2\pi)^2\varphi_3^2/2L
=2E_J-2E_J\cos\theta\cos(\varphi-\pi f)+2(\Phi_0/2\pi)^2\varphi^2/L$. Conjugate to $\theta$ and $\varphi$, the canonical momenta are $P_{\theta}=\frac{\partial{\cal L}}{\partial{\dot\theta}}=2(\Phi_0/2\pi)^2(C_s+C_J){\dot\theta}$, and
$P_{\varphi}=\frac{\partial{\cal L}}{\partial{\dot\varphi}}=2(\Phi_0/2\pi)^2(C_s+C_J+2C_L){\dot\varphi}$.
Then, the Hamiltonian of the pokemon, $H_q=P_{\theta}{\dot\theta}^2+P_{\varphi}{\dot\varphi}^2-{\cal L}$, can be expressed as
\begin{equation}\label{H1}
H_q=\frac{P_{\theta}^2}{4(\Phi_0/2\pi)^2(C_s+C_J)}+\frac{P_{\varphi}^2}{4(\Phi_0/2\pi)^2(C_s+C_J+2C_L)}+U,
\end{equation}
with the commutation relations $[\theta,P_{\theta}]=i\hbar$, and $[\varphi,P_{\varphi}]=i\hbar$. When introducing number operators for Cooper pairs, $\hat{N}_{\theta}=-i\frac{\partial}{\partial\theta}$ and $\hat{N}_{\varphi}=-i\frac{\partial}{\partial\varphi}$, Hamiltonian (\ref{H1}) is converted to
\begin{equation}\label{Hq}
H_q=H_J+E_L\varphi^2,
\end{equation}
with the inductive energy $E_L=2(\Phi_0/2\pi)^2/L$, and
\begin{equation}\label{HJ}
H_J=4E_{c\theta}\hat{N}_{\theta}^2+4E_{c\varphi}\hat{N}_{\varphi}^2+2E_J-2E_J\cos\theta\cos(\varphi-\pi f).
\end{equation}
Here $E_{c\theta}=e^2[4(C_s+C_J)]^{-1}\equiv \tilde{E}_{c\theta}$ and $E_{c\varphi}=e^2[4(C_s+C_J+2C_L)]^{-1}\equiv\tilde{E}_{c\varphi}$ are single-electron charging energies relevant to the two junctions.

For the 0-$\pi$ qubit in Fig.~\ref{fig1}(a), the phase drops across the two Josephson junctions and the two inductors are constrained by $\varphi_1-\varphi_2-\varphi_3+\varphi_4+2\pi f=0$. The Hamiltonian can be written as~\cite{Dempster-PRA-14} $H_{\rm tot}=H_q+H_{\rm osc}$, where $H_q$ has the same form as in Eq.~(\ref{Hq}), but $E_{c\theta}=\tilde{E}_{c\theta}$, $E_{c\varphi}=e^2[4C_J]^{-1}$, and $E_L=(\Phi_0/2\pi)^2/L$. Here $H_{\rm osc}=4E_{c\chi}\hat{N}_{\chi}^2+E_L\chi^2$, with $E_{c\chi}=e^2[4C_s]^{-1}$ and $\hat{N}_{\chi}=-i\frac{\partial}{\partial\chi}$, is the Hamiltonian of a harmonic oscillator. The canonical coordinates conjugate to the number operators $\hat{N}_{\theta}$, $\hat{N}_{\varphi}$ and $\hat{N}_{\chi}$ are $\theta\equiv\frac{1}{2}(\varphi_1+\varphi_2)$, $\varphi\equiv\frac{1}{2}(\varphi_3-\varphi_4)$, and
$\chi\equiv\frac{1}{2}(\varphi_3+\varphi_4)$, respectively. Since $H_{\rm osc}$ is fully decoupled from $H_q$, the Hamiltonian of the 0-$\pi$ qubit can be reduced to $H_q$.

\begin{figure}
\includegraphics[width=0.48\textwidth]{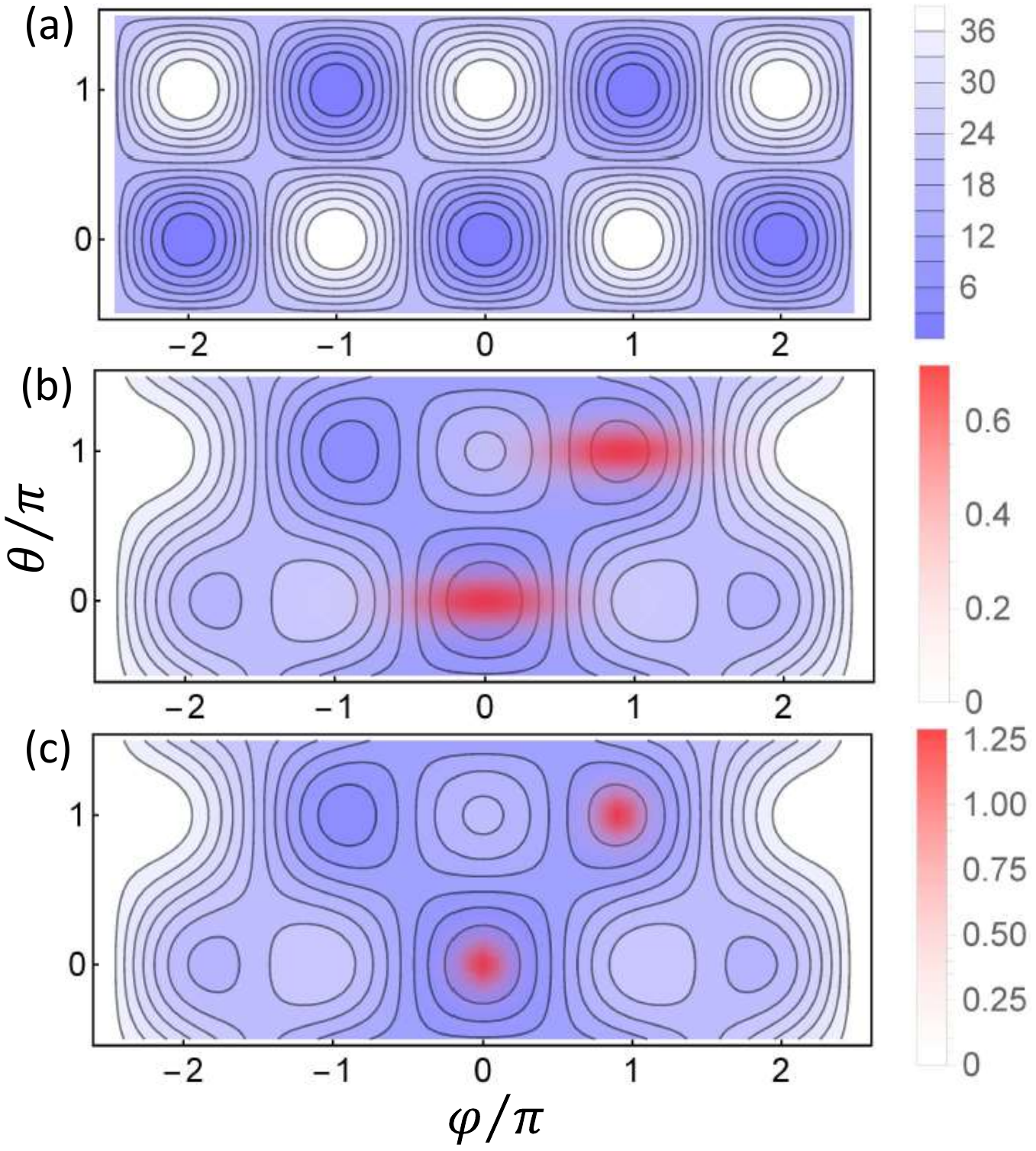}
\caption{(a)~Contour plot of the potential $U_J=2E_J-2E_J\cos\theta\cos(\varphi-\pi f)$ at $f=0$. (b) and (c):~Wave functions of the two basis states $|0\rangle$ and $|1\rangle$ for (b) the 0-$\pi$ qubit and (c) the pokemon, where the contour plot corresponds to the potential $U=U_J+E_L\varphi^2$. The parameters are set to be $E_{c\theta}=0.1$, $E_J=10$, and $E_L=1$, but $E_{c\varphi}=10$ in (b), and $E_{c\varphi}=0.09$ in (c).}
\label{fig2}
\end{figure}

For both the 0-$\pi$ qubit in Fig.~\ref{fig1}(a) and the pokemon in Fig.~\ref{fig1}(b), when the external magnetic flux is absent (i.e., $f=0$), the potential without the inductive energy, i.e., $U_J=2E_J-2E_J\cos\theta\cos(\varphi-\pi f)$, has degenerate minima at points $(\theta,\varphi)=(0,0)$ and ($\pi$,$\pi$) in the 2D parameter space [see Fig.~\ref{fig2}(a)]. Nevertheless, the inductive potential $E_L\varphi^2$ removes this degeneracy [see the contour plot in Figs.~\ref{fig2}(b) and \ref{fig2}(c)]. When $E_J\gg E_L$, as in the 0-$\pi$ qubit~\cite{Brooks-PRA-13}, we can use the lowest-energy states localized around (0,0) and ($\pi$,$\pi$) to encode the pokemon qubit. Therefore, both the pokemon and the 0-$\pi$ qubit are actually logic qubits encoded on {\it two} degrees of freedom $\theta$ and $\varphi$. However, there are distinct differences between them. First, the pokemon has a simpler circuitry, providing an advantage in sample fabrication. Second, the pokemon has two {\it small} single-electron charging energies $E_{c\theta}=\tilde{E}_{c\theta}$ and $E_{c\varphi}=\tilde{E}_{c\varphi}$, because $C_s\gg C_J$. This corresponds to a {\it heavy} particle with anisotropic masses $M_{\theta}=2(\Phi_0/2\pi)^2(C_s+C_J)$, and $M_{\varphi}=2(\Phi_0/2\pi)^2(C_s+C_J+2C_L)$, moving in quantum wells separated by a barrier of height $E_J\gg E_{c\theta},E_{c\varphi}$. Owing to these large masses in both $\theta$ and $\varphi$ directions as well as the higher inter-well barrier, the ground and first excited states $|0\rangle$ and $|1\rangle$ of the pokemon are highly localized in the vicinity of (0,0) and ($\pi$,$\pi$), respectively, along  both $\theta$ and $\varphi$ directions [Fig.~\ref{fig2}(c)]; and the transition frequency of the pokemon qubit is mainly determined by the inductive-potential difference between these two points. As for the 0-$\pi$ qubit, the effective mass along the $\theta$ direction is as large as $M_{\theta}$ of the pokemon, but the effective mass along the $\varphi$ direction, $M_{\varphi}=2(\Phi_0/2\pi)^2C_J$, is much smaller. The ground and first excited states of the 0-$\pi$ qubit are hence only highly localized at (0,0) and ($\pi$,$\pi$) along the $\theta$ direction [Fig.~\ref{fig2}(b)]. This indicates that the pokemon can become more protected.

In addition, when the capacitance $C_L$ is absent in Fig.~\ref{fig1}(b), the pokemon is reduced to the circuit configuration of the bifluxon~\cite{bifluxon-20}, but without the gate, and each Josephson junction there is now shunted by a large capacitance $C_s$. This capacitively shunted bifluxon has the same Hamiltonian as in Eq.~(\ref{Hq}), but with $E_{c\theta}=E_{c\varphi}=\tilde{E}_{c\theta}$, i.e., the effective masses become isotropic.

{\it Robustness against both charge and flux noises.}{\bf---}Below we reveal the robustness of the pokemon against both charge and flux noises, which are usually two major decoherence sources in superconducting circuits. For the 0-$\pi$ qubit, since the ground and first excited states $|0\rangle$ and $|1\rangle$ are less localized around (0,0) and ($\pi$,$\pi$) along the $\varphi$ direction, it is suitable to numerically study its quantum coherence~\cite{Dempster-PRA-14}. On the contrary, the ground and first excited states $|0\rangle$ and $|1\rangle$ of the pokemon are highly localized around (0,0) and ($\pi$,$\pi$) along both $\theta$ and $\varphi$ directions. These two basis states of the pokemon can be well approximated by
$|0\rangle=\sqrt{\frac{\alpha_{\theta}\alpha_{\varphi}}{\pi}}
e^{-\frac{1}{2}(\alpha_{\theta}^2\theta^2+\alpha_{\varphi}^2\varphi^2)}$, and
$|1\rangle=\sqrt{\frac{\alpha_{\theta}\alpha_{\varphi}}{\pi}}e^{-\frac{1}{2}[\alpha_{\theta}^2(\theta-\pi)^2
+\alpha_{\varphi}^2(\varphi-\frac{E_J}{E_J+E_L}\pi)^2]}$,
with $\alpha_{\theta}^2=\frac{1}{2}\sqrt{\frac{E_J}{E_{c\theta}}}$, and $\alpha_{\varphi}^2=\frac{1}{2}\sqrt{\frac{E_J+E_L}{E_{c\varphi}}}$.
Due to the inductive potential, the center of $|1\rangle$ is shifted to $\frac{E_J}{E_J+E_L}\pi$ in the $\varphi$ direction, and the level difference between $|1\rangle$ and $|0\rangle$ is $\varepsilon_{10}=\frac{E_JE_L\pi^2}{E_J+E_L}\sim E_L\pi^2$ for $E_J\gg E_L$.

When both charge and flux fluctuations are considered, the Hamiltonian (\ref{Hq}) is changed, at $f=0$, to
\begin{eqnarray}\label{Ht}
H_{t}&=&4E_{c\theta}(\hat{N}_{\theta}-\delta N_{\theta})^2+4E_{c\varphi}(\hat{N}_{\varphi}-\delta N_{\varphi})^2 \nonumber\\
&&+2E_J-2E_J\cos\theta\cos(\varphi-\pi\delta\!f)+E_L\varphi^2,
\end{eqnarray}
where $\delta N_{\theta(\varphi)}\equiv\delta Q_{\theta(\varphi)}/2e$ is the reduced charge fluctuation and $\delta\!f
\equiv\delta\Phi_{\rm ext}/\Phi_0$ is the reduced flux fluctuation.
Up to second-order perturbations, the Hamiltonian (\ref{Ht}) can be written as $H_t=H_q+H^{\prime}_{\theta}+H^{\prime}_{\varphi}+H^{\prime}_f$, with $H_q$ given by Eq.~(\ref{Hq}) and
\begin{eqnarray}
H^{\prime}_{\theta(\varphi)}&=&X_{\theta(\varphi)}^{(1)}\delta N_{\theta(\varphi)}+\frac{1}{2}X_{\theta(\varphi)}^{(2)}\delta N^2_{\theta(\varphi)}, \nonumber\\
H^{\prime}_f&=&X_f^{(1)}\delta\!f+\frac{1}{2}X_f^{(2)}\delta\!f^2,
\end{eqnarray}
where $X_{\theta(\varphi)}^{(1)}\equiv (\frac{\partial H_t}{\partial\,\delta N_{\theta(\varphi)}})|_{\delta N_{\theta(\varphi)}=0}=-8E_{c\theta(\varphi)}\hat{N}_{\theta(\varphi)}$,
$X_{\theta(\varphi)}^{(2)}\equiv (\frac{\partial^2 H_t}{\partial\,\delta N^2_{\theta(\varphi)}})|_{\delta N_{\theta(\varphi)}=0}=8E_{c\theta(\varphi)}$,
$X_f^{(1)}\equiv (\frac{\partial H_t}{\partial\,\delta\!f})|_{\delta\!f=0}=-2\pi E_J\cos\theta\sin\varphi$, and
$X_f^{(2)}\equiv (\frac{\partial^2 H_t}{\partial\,\delta\!f^2})|_{\delta\!f=0}=2\pi E_J\cos\theta\cos\varphi$.

Note that the perturbation arising from the charge noise terminates at the second order owing to the specific form of the Hamiltonian (\ref{Ht}). We can derive that $\langle 1|X_{\theta(\varphi)}^{(1)}|1\rangle-\langle 0|X_{\theta(\varphi)}^{(1)}|0\rangle=0$, and $\langle 1|X_{\theta(\varphi)}^{(2)}|1\rangle-\langle 0|X_{\theta(\varphi)}^{(2)}|0\rangle=0$. These imply that the charge fluctuation does not induce dephasing to the pokemon.
However, for the flux noise, we have nonzero $A_f\equiv\frac{1}{2}(\langle 1|X_f^{(1)}|1\rangle-\langle 0|X_f^{(1)}|0\rangle)$:
\begin{equation}\label{Af}
A_f=2\pi E_J\sin\left(\frac{E_L\pi}{E_J+E_L}\right)
\exp\left\{-\frac{1}{2}\left(\sqrt{\frac{E_{c\theta}}{E_J}}+\sqrt{\frac{E_{c\varphi}}{E_J+E_L}}\right)\right\},
\end{equation}
which yields dephasing to the pokemon. The corresponding dephasing-induced decay factor is $e^{-\eta(t)}$, with~\cite{supplemental}
\begin{equation}\label{dephasing}
\eta(t)=\frac{1}{\hbar^2}|A_f|^2\!\!\int^{+\infty}_{\omega_c}\!\!d\omega\; S_{\!f}(\omega)\frac{\sin^2(\omega t/2)}{2\pi(\omega/2)^2},
\end{equation}
where $\omega_c$ is a low-frequency cutoff of the flux-noise power spectrum $S_{\!f}(\omega)=\int^{+\infty}_{-\infty}dt \langle \delta\!f(t'+t)\delta\!f(t')\rangle e^{-i\omega t}$. The dephasing rate can be defined as $\Gamma_{\phi}=1/T_{\phi}$, with the dephasing time $T_{\phi}$ determined by $\eta(T_{\phi})=1$.

Moreover, we can derive that $\langle 0|X_f^{(1)}|1\rangle=0$ and $\langle 0|X_f^{(2)}|1\rangle=0$ for the flux noise. Similar to $X_f^{(1)}$ and $X_f^{(2)}$, higher-order perturbations arising from flux noise also contain either $\cos\theta\sin\varphi$ or $\cos\theta\cos\varphi$, so $\langle 0|X_f^{(n)}|1\rangle=0$ for $n\ge 3$. Hence the flux noise does not induce relaxation to the pokemon. Instead, the first-order charge fluctuations yield relaxation to the pokemon, because
\begin{eqnarray}\label{Bfactor}
B_{\theta}\equiv\langle 0|X_{c\theta}^{(1)}|1\rangle &=& i2\pi \sqrt{E_{c\theta}E_J}\;e^{-\gamma},   \nonumber\\
B_{\varphi}\equiv\langle 0|X_{c\varphi}^{(1)}|1\rangle &=& i2\pi \sqrt{E_{c\varphi}(E_J+E_L)}\;e^{-\gamma},
\end{eqnarray}
where the decay rate $\gamma$ is
\begin{equation}\label{gamma}
\gamma=\frac{\pi^2}{8}\left[\sqrt{\frac{E_J}{E_{c\theta}}}
+\left(\frac{E_J}{E_J+E_L}\right)^2\sqrt{\frac{E_J+E_L}{E_{c\varphi}}}\right].
\end{equation}
According to the Fermi golden rule~\cite{Ithier-PRB-05}, the relaxation rate for each charge noise can be obtained as
\begin{equation}\label{relax}
\Gamma_{1,i}=\frac{1}{\hbar^2}|B_{i}|^2S_{\!i}(\omega_{10}),
\end{equation}
where each charge-noise power spectrum is defined as $S_{\!i}(\omega)=\int^{+\infty}_{-\infty}dt \langle \delta N_{i}(t'+t)\delta N_{i}(t')\rangle e^{-i\omega t}$, with $i=\theta$, $\varphi$, and $\omega_{10}=\varepsilon_{10}/\hbar$ is the transition frequency of the pokemon. The total relaxation rate is $\Gamma_1=\Gamma_{1,\theta}+\Gamma_{1,\varphi}$, which gives the relaxation time $T_1=1/\Gamma_1$ of the pokemon.

For superconducting circuits, low-frequency noise plays a pivotal role in decoherence, which can be modeled as $1/f$ noise via $S_i(\omega)=K_i/|\omega|$, where $i=\theta$, $\varphi$, and $f$. Typically, $K_{\theta(\varphi)}\sim 1.7\times 10^{-6}$ for  charge noise~\cite{Nakamura-PRL-02} and $K_{f}\sim 3\times 10^{-12}$ for flux noise~\cite{Mooij-PRL-05}.
For $E_{c\theta(\varphi)},E_L\ll E_J$, $A_f$ reduces to a small quantity $A_f\sim 2\pi^2E_L$ [see Eq.~(\ref{Af})], in comparison with $E_J$. Also, because $K_f$ is several orders of magnitude smaller than $K_{\theta(\varphi)}$, a long dephasing time $T_{\phi}$ is given to satisfy $\eta(T_{\phi})=1$ in Eq.~(\ref{dephasing}), so it gives a small dephasing rate $\Gamma_{\phi}$. Then, we can ignore this dephasing rate and write the decoherence rate as $\Gamma_2=\frac{1}{2}\Gamma_1+\Gamma_{\phi}\approx \frac{1}{2}\Gamma_1$. Moreover, it follows from Eqs.~(\ref{Bfactor})-(\ref{relax}) that the relaxation rate is proportional to $e^{-2\gamma}$, which is {\it exponentially} reduced by both $\sqrt{\frac{E_J}{E_{c\theta}}}$ and $\sqrt{\frac{E_J+E_L}{E_{c\varphi}}}$. Because $E_{c\theta(\varphi)},E_L\ll E_J$, these two quantities are both large, yielding a small relaxation rate as well. Therefore, the pokemon is a well-protected logic qubit, owing to its {\it small} rates in {\it both} dephasing and relaxation. In contrast, the 0-$\pi$ qubit has $E_{c\theta}\ll E_J$ but not $E_{c\varphi}\ll E_J$, so its relaxation rate is only exponentially reduced by $\sqrt{\frac{E_J}{E_{c\theta}}}$, without a further exponential reduction by $\sqrt{\frac{E_J+E_L}{E_{c\varphi}}}\approx\sqrt{\frac{E_J}{E_{c\varphi}}}$ in the pokemon.

\begin{figure}
\includegraphics[width=0.48\textwidth]{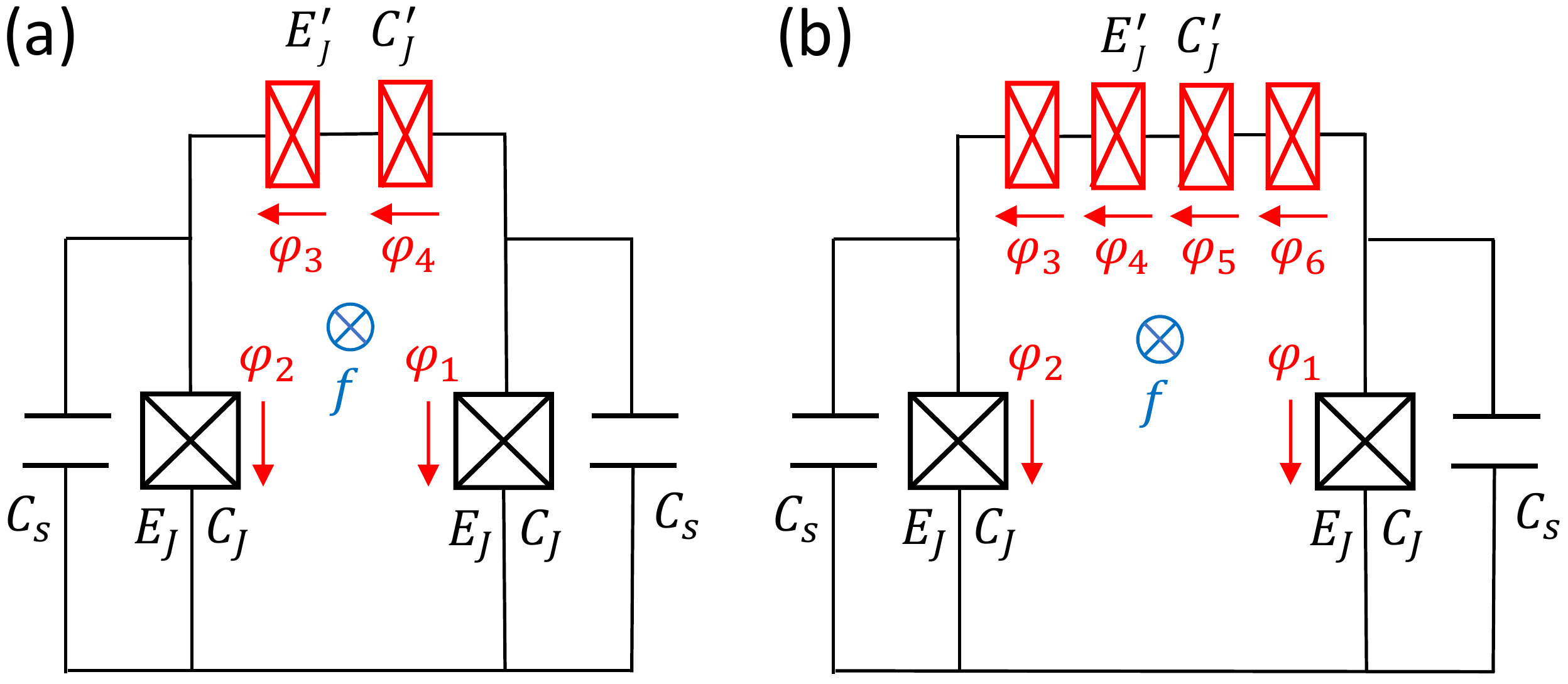}
\caption{(a)~The pokemon modified by replacing the capacitively shunted inductor in Fig.~\ref{fig1}(b) with two (i.e., a pair of) Josephson junctions, both having the same coupling energy $E^{\prime}_J$ and capacitance $C^{\prime}_J$.
(b) ~The pokemon modified by replacing the capacitively shunted inductor with two pairs of Josephson junctions, where each junction has the same coupling energy $E^{\prime}_J$ and capacitance $C^{\prime}_J$.
Extension of the pokemon applies when modified by replacing the capacitively shunted inductor with more Josephson junctions.}
\label{fig3}
\end{figure}

{\it Pokemon with the loop inductance replaced by a nonlinear inductance.}{\bf---}For both the 0-$\pi$ qubit in Fig.~\ref{fig1}(a) and pokemon in Fig.~\ref{fig1}(b), it is required that $E_J\gg E_L$, needing a large loop inductance $L$. Here we propose to implement the pokemon by replacing the loop inductance with a nonlinear one.

We first replace the capacitively shunted inductor in Fig.~\ref{fig1}(b) with two Josephson junctions, both having the same coupling energy $E^{\prime}_J$ and capacitance $C^{\prime}_J$ [see Fig.~\ref{fig3}(a)]. The total Hamiltonian of the pokemon can be written as~\cite{supplemental}
\begin{equation}\label{Htot}
H_{\rm tot}=H_J+2E_{J}^{\prime}(1-\cos\varphi)\cos\chi+H_{\chi},
\end{equation}
where $H_J$ has the same form as in Eq.~(\ref{HJ}), but with $E_{c\theta}=\tilde{E}_{c\theta}$, and $E_{c\varphi}=e^2[4(C_s+C_J+C_J^{\prime})]^{-1}$; $H_{\chi}=4E_{c\chi}N_{\chi}^2+2E_J^{\prime}(1-\cos\chi)$, with $E_{c\chi}=e^2[4C_J^{\prime}]^{-1}$. For the subsystem with $H_{\chi}$, the corresponding harmonic-oscillator frequency is $\omega_{\chi}=\frac{4}{\hbar}\sqrt{E_J^{\prime}E_{c\chi}}$. By suitably choosing parameters to have a sufficiently large $\omega_{\chi}$, the system stays in the ground state $|g\rangle$ of $H_{\chi}$, and the Hamiltonian (\ref{Htot}) can be reduced to
%\begin{equation}\label{Hq-1}
$H_q=H_J+E^{(1)}_J(1-\cos\varphi)$,
%\end{equation}
where $E^{(1)}_J=2E^{\prime}_J\langle g|\cos\chi|g\rangle$. When $E_{c\chi}\ll E'_J$, we have $E^{(1)}_J=2E^{\prime}_J e^{-\eta}$, with $\eta=\frac{1}{2}\sqrt{\frac{E_{c\chi}}{E'_J}}$. Comparing with Eq.~(\ref{Hq}), the last term in $H_q$, i.e., $E^{(1)}_J(1-\cos\varphi)$, is the inductive potential of a nonlinear inductor, which also removes the degeneracy of the basis states $|0\rangle$ and $|1\rangle$ at $(0,0)$ and $(\pi,\pi)$. The energy-level difference between these two basis states of the pokemon is $\varepsilon_{10}=2E^{(1)}_J\sim 4E'_J$. To avoid any unwanted levels in between $|0\rangle$ and $|1\rangle$, we can take $\omega_{\chi}>\omega_q\equiv\varepsilon_{10}/\hbar$.

When replacing the capacitively shunted inductor with four (i.e., two pairs of) Josephson junctions, each junction having coupling energy $E^{\prime}_J$ and capacitance $C^{\prime}_J$ [see Fig.~\ref{fig3}(b)], the Hamiltonian of the pokemon can be reduced to~\cite{supplemental}
%\begin{equation}\label{Hq-2}
$H_q=H_J+E^{(2)}_J[1-\cos(\varphi/2)]$.
%\end{equation}
When $E'_J\gg E_{c\xi}\equiv e^2[4C'_J]^{-1}$, we have $E^{(2)}_J=4E'_J e^{-(\eta+\eta')}$, with $\eta=\frac{1}{2}\sqrt{\frac{E_{c\xi}}{E'_J}}$, and $\eta'=\frac{1}{2}\eta e^{\eta/2}$. Also, $H_J$ has the same form as in Eq.~(\ref{HJ}), but with $E_{c\theta}=\tilde{E}_{c\theta}$, and $E_{c\varphi}=e^2[4(C_s+C_J+C_J^{\prime}/2)]^{-1}$.
Similarly, the last term in $H_q$, i.e., $E^{(2)}_J[1-\cos(\varphi/2)]$, acts as the inductive potential of a nonlinear inductor, which yields an energy-level difference between the two basis states $|0\rangle$ and $|1\rangle$ of the pokemon, $\varepsilon_{10}=E^{(2)}_J\sim 4E'_J$.
Here we have shown that a small number of Josephson junctions can be used to achieve the inductive potential of a nonlinear inductor to remove the state degeneracy around (0,0) and ($\pi$,$\pi$) for the two basis states of the pokemon. This avoids the complex circuitry in the experimentally realized 0-$\pi$ qubit~\cite{Gyenis-PRXQ-21} where a large number of Josephson junctions were used to implement a linear inductor.

{\it Conclusions.}{\bf---}In summary, a new protected logic qubit called the pokemon is proposed by harnessing two capacitively shunted Josephson junctions and one capacitively shunted inductor embedded in a superconducting loop. Similar to the 0-$\pi$ qubit, the two basis states of the proposed qubit are also separated by a high barrier, but their wave functions can be highly localized along {\it both} axis directions of the 2D parameter space, in sharp contrast to the highly localized wave functions along only one axis direction in the 0-$\pi$ qubit. This makes the pokemon qubit more protected. Here we find that the flux noise does not induce relaxation but weak dephasing to the pokemon, while the charge noise does not cause dephasing and the induced relaxation can be exponentially reduced by the large capacitors shunted to the two Josephson junctions.
Furthermore, we show how to achieve the pokemon by replacing the capacitively shunted inductor with a few (one or two) pairs of Josephson junctions. This offers an experimentally feasible method to implement protected logic qubits with even higher quantum coherence.
In the near future, an experimental comparison between the properties of a pokemon and a transmon would be insightful. It would be the superconducting qubit analog of Godzilla meets King Kong.

\section*{Acknowledgments}
We thank Guo-Qiang Zhang for technical assistance. J.Q.Y. is supported by the National Natural Science Foundation of China (Grants No. 11934010 and No. U1801661), the National Key Research and Development Program of China (Grant No. 2016YFA0301200), and the Zhejiang Province Program for Science and Technology (Grant No. 2020C01019). F.N. is supported in part by Nippon Telegraph and Telephone Corporation (NTT) Research, the Army Research Office (ARO) (Grant No. W911NF-18-1-0358), the Japan Science and Technology Agency (JST) [via the Quantum Leap Flagship Program (Q-LEAP) program and the Centers of Research Excellence in Science and Technology (CREST) Grant No. JPMJCR1676], the Japan Society for the Promotion of Science (JSPS) (via the Grants-in-Aid for Scientific Research (KAKENHI) Grant No. JP20H00134 and the Japan Society for the Promotion of Science (JSPS)-Russian Foundation for Basic Research (RFBR) Grant No. JPJSBP120194828), the Asian Office of Aerospace Research and Development (AOARD) (via Grant No. FA2386-20-1-4069), and the Foundational Questions Institute Fund (FQXi) via Grant No. FQXi-IAF19-06.


\begin{thebibliography}{99}

\bibitem{Nori-RMP-14}
I. M. Georgescu, S. Ashhab, and F. Nori, Quantum simulation, Rev. Mod. Phys. {\bf 86}, 153 (2014).

\bibitem{Aspuru-Guzik-RMP-20}
S. McArdle, S. Endo, A. Aspuru-Guzik, S. C. Benjamin, and X. Yuan, Quantum computational chemistry,
Rev. Mod. Phys. {\bf 92}, 015003 (2020).

\bibitem{Nielsen-Chuang}
M. A. Nielsen and I. Chuang, {\it Quantum Computation and Quantum Information} (Cambridge Univ. Press, 2000).

\bibitem{Clarke-PRL-85}
J. M. Martinis, M. H. Devoret, J. Clarke, Energy-Level Quantization in the Zero-Voltage State of a Current-Biased Josephson Junction, Phys. Rev. Lett. {\bf 55}, 1543 (1985).

\bibitem{Devoret-PScip-98}
V. Bouchiat, D. Vion, P. Joyez, D. Esteve, and M. H. Devoret, Quantum coherence with a single Cooper pair,
Phys. Scripta {\bf T76}, 165 (1998).

\bibitem{Nakamura-Nature-1999}
Y. Nakamura, Y. A. Pashkin, J. S. Tsai, Coherent control of macroscopic quantum states in a single-Cooper-pair box.
Nature {\bf 398}, 786 (1999).

\bibitem{Mooij-Sci-99}
J. E. Mooij, T. P. Orlando, L. Levitov, L. Tian, C. H. van der Waland, and S. Lloyd, Josephson Persistent-Current Qubit,
Science {\bf 285}, 1036 (1999).

\bibitem{Vion-Science-02}
D. Vion, A. Aassime, A. Cottet, P. Joyez, H. Pothier, C. Urbina, D. Esteve, and M. H. Devoret,
Manipulating the quantum state of an electrical circuit, Science {\bf 296},
886 (2002).

\bibitem{Yu-Sci-02}
Y. Yu, S. Han, X. Chu, S.-I. Chu, and Z. Wang, Coherent Temporal Oscillations of Macroscopic Quantum States in a Josephson Junction, Science {\bf 296}, 889 (2002).

\bibitem{Martinis-PRL-02}
J. M. Martinis, S. Nam, J. Aumentado, and C. Urbina, Rabi oscillations in a large
Josephson-junction qubit. Phys. Rev. Lett. {\bf 89}, 117901 (2002).

\bibitem{Nori-PhysRep-17}
X. Gu, A. F. Kockum, A. Miranowicz, Y. X. Liu, and F. Nori, Microwave photonics with superconducting quantum circuits,
Phys. Rep. {\bf 718-719}, 1 (2017).

\bibitem{Martinis-Nature-19}
F. Arute {\it et al.}, Quantum supremacy using a programmable superconducting processor,
Nature {\bf 574}, 505 (2019).

\bibitem{Zhu-PRL-21}
Y. Wu {\it et al.}, Strong Quantum Computational Advantage Using a Superconducting Quantum Processor,
Phys. Rev. Lett. {\bf 127}, 180501 (2021).

\bibitem{Siddiqi-NatRev-21}
I. Siddiqi, Engineering high-coherence superconducting qubits,
Nat. Rev. Materials {\bf 6}, 875 (2021).

\bibitem{Tsai-JAP-21}
S. Kwon, A. Tomonaga, G. L. Bhai, S. J. Devitt, and J. S. Tsai, Gate-based superconducting quantum computing,
J. Appl. Phys. {\bf 129}, 041102 (2021).

\bibitem{You-PRB-07}
J. Q. You, X. Hu, S. Ashhab, and F. Nori, Low-decoherence flux qubit, Phys. Rev. B {\bf 75}, 140515(R) (2007).

\bibitem{Steffen-PRL-10}
M. Steffen, S. Kumar, D. P. DiVincenzo, J. R. Rozen, G. A. Keefe, M. B. Rothwell, and M. B. Ketchen,
High-Coherence Hybrid Superconducting Qubit, Phys. Rev. Lett. {\bf 105}, 100502 (2010).

\bibitem{Yan-NC-16}
F. Yan, S. Gustavsson, A. Kamal, J. Birenbaum, A. P. Sears, D. Hover, T. J. Gudmundsen, D. Rosenberg,
G. Samach, S. Weber, J. L. Yoder, T. P. Orlando, J. Clarke, A. J. Kerman, and W. D. Oliver,
The flux qubit revisited to enhance coherence and reproducibility, Nat. Commun. {\bf 7}, 12964 (2016).

\bibitem{Koch-PRA-07}
J. Koch, T. M. Yu, J. Gambetta, A. A. Houck, D. I. Schuster, J. Majer, A. Blais, M. H. Devoret, S. M. Girvin, and R. J. Schoelkopf, Charge-insensitive qubit design derived from the Cooper pair box.
Phys. Rev. A {\bf 76}, 042319 (2007).

\bibitem{Martinis-PRL-13}
R. Barends, J. Kelly, A. Megrant, D. Sank, E. Jeffrey, Y. Chen, Y. Yin, B. Chiaro, J. Mutus, C. Neill, P. O¡¯Malley,
P. Roushan, J. Wenner, T. C. White, A. N. Cleland, and John M. Martinis,
Coherent Josephson Qubit Suitable for Scalable Quantum Integrated Circuits, Phys. Rev. Lett. {\bf 111}, 080502 (2013).

\bibitem{Larsen-PRL-15}
T. W. Larsen, K. D. Petersson, F. Kuemmeth, T. S. Jespersen, P. Krogstrup, J. Nyg{\aa}rd, and C. M. Marcus,
Semiconductor-Nanowire-Based Superconducting Qubit, Phys. Rev. Lett. {\bf 115}, 127001 (2015).

\bibitem{Earnest-PRL-18}
N. Earnest, S. Chakram, Y. Lu, N. Irons, R. K. Naik, N. Leung, L. Ocola, D. A. Czaplewski, B. Baker, J. Lawrence, J. Koch,
and D. I. Schuster,
Realization of a $\Lambda$ System with Metastable States of a Capacitively Shunted Fluxonium,
Phys. Rev. Lett. {\bf 120}, 150504 (2018).

\bibitem{Wendin-Review-17}
G. Wendin, Quantum information processing with superconducting circuits: a review.
Rep. Prog. Phys. {\bf 80}, 106001 (2017).

\bibitem{Preskill-Quantum-18}
J. Preskill, Quantum computing in the NISQ era and beyond. Quantum {\bf 2}, 79 (2018)

%\bibitem{Ioffe-Review-12}
%B. Dou\c{c}ot, B. and L. B. Ioffe, Physical implementation of protected qubits, Rep. Prog. Phys. {\bf 75}, 072001 (2012).

\bibitem{Brooks-PRA-13}
P. Brooks, A. Kitaev, and J. Preskill, Protected gates for superconducting qubits,
Phys. Rev. A {\bf 87}, 052306 (2013).

\bibitem{Gyenis-PRXQ-21}
A. Gyenis, P. S. Mundada, A. Di Paolo, T. M. Hazard, X. You, D. I. Schuster, J. Koch, A. Blais, and A. A. Houck,
Experimental Realization of a Protected Superconducting Circuit Derived from the 0-$\pi$ Qubit,
PRX Quantum 2, 010339 (2021).

%\bibitem{Manucharyan-Sci-09}
%V. E. Manucharyan, J. Koch, L. I. Glazman, and M. H. Devoret, Fluxonium: Single cooper-pair circuit free of charge offsets,
%Science {\bf 326}, 113 (2009).

\bibitem{bifluxon-20}
K. Kalashnikov, W. T. Hsieh, W. Zhang, W.-S. Lu, P. Kamenov, A. Di Paolo, A. Blais, M. E. Gershenson, and M. Bell,
Bifluxon: Fluxon-Parity-Protected Superconducting Qubit, PRX Quantum {\bf 1}, 010307 (2020).

\bibitem{Dempster-PRA-14}
J. M. Dempster, B. Fu, D. G. Ferguson, D. I. Schuster, and J. Koch, Understanding degenerate ground states
of a protected quantum circuit in the presence of disorder,
Phys. Rev. B {\bf 90}, 094518 (2014).

\bibitem{supplemental}
See Supplemental Material.

\bibitem{Ithier-PRB-05}
G. Ithier, E. Collin, P. Joyez, P. J. Meeson, D. Vion, D. Esteve, F. Chiarello, A. Shnirman, Y. Makhlin,
J. Schriefl, and G. Sch{\" o}n,
Decoherence in a superconducting quantum bit circuit, Phys. Rev. B {\bf 72}, 134519 (2005).

\bibitem{Nakamura-PRL-02}
Y. Nakamura, Yu. A. Pashkin, T. Yamamoto, and J. S. Tsai, Charge Echo in a Cooper-Pair Box, Phys. Rev. Lett. {\bf88}, 047901 (2002).

\bibitem{Mooij-PRL-05}
P. Bertet, I. Chiorescu, G. Burkard, K. Semba, C. J. P. M. Harmans, D. P. DiVincenzo, and J. E. Mooij,
Dephasing of a Superconducting Qubit Induced by Photon Noise, Phys. Rev. Lett. {\bf 95}, 257002 (2005).



\end{thebibliography}
\end{document}